# Anomalies in thickness measurements of graphene and few layer graphite crystals by tapping mode atomic force microscopy

P. Nemes-Incze[1, *], Z. Osváth[1], K. Kamarás[2], L.P. Biró[1]

*1 Research Institute for Technical Physics and Materials Science, Hungarian Academy of Sciences, H-1525 Budapest, P.O. Box 49, Hungary*
*2 Research Institute for Solid State Physics and Optics, Hungarian Academy of Sciences, H-1525 Budapest, P.O. Box 49, Hungary*

## Abstract

Atomic Force Microscopy (AFM) in the tapping (intermittent contact) mode is a commonly used tool to measure the thickness of graphene and few layer graphene (FLG) flakes on silicon oxide surfaces. It is a convenient tool to quickly determine the thickness of individual FLG films. However, reports from literature show a large variation of the measured thickness of graphene layers. This paper is focused on the imaging mechanism of tapping mode AFM (TAFM) when measuring graphene and FLG thickness and we show that at certain measurement parameters significant deviations can be introduced in the measured thickness of FLG flakes. An increase of as much as 1 nm can be observed in the measured height of FLG crystallites, when using an improperly chosen range of free amplitude values of the tapping cantilever. We present comparative Raman spectroscopy and TAFM measurements on selected single and multilayer graphene films, based on which we suggest ways to correctly measure graphene and FLG thickness using TAFM.

## Introduction

Graphene, as a the building block of graphite has been theoretically investigated since the 1940s [1]. Until 2004, when Novoselov et al. successfully identified graphene [2] and other 2D crystals [3] in a simple tabletop experiment, it was assumed that 2D crystals were thermodynamically unstable and could not exist under ambient conditions [4, 5]. The discovery that such samples can be produced has led to a wealth of scientific investigation [6, 7, 8, 9, 10, 11], due to the very promising electronic [12] and mechanical properties of graphene, furthermore due to it's resistance to mechanical and chemical stress and it's high crystallinity [3].

Today, the most successful method to prepare graphene samples is mechanical exfoliation of graphite onto oxidised Si wafers [2]. The thickness of few layer graphite (FLG) films is estimated by optical microscopy [13, 14], after which the thickness of the thinnest crystallites is measured by AFM. Much like other scanning probe techniques, AFM is not free of measurement artefacts. With AFM being such a widely employed tool to inspect the thickness of FLG crystals, we believe that a detailed investigation of the possible sources of errors in the AFM measurement of FLG thickness is of great importance. When measuring features of the order of magnitude of one atomic layer and changing material properties, effects of lesser importance under other circumstances (like homogeneous samples), may play a crucial role in distorting topography images. Indeed various groups reported different thickness measurements for graphene layers, with thicknesses ranging from 0.35 nm to 1 nm, relative to the $SiO_2$ substrate. Novoselov et al. measured platelet thicknesses of 1-1.6 nm [2]. Gupta et al. have measured an instrumental offset induced by the AFM, of 0.33 nm, ie. 0.7 nm height for a single layer [20]. Other authors have also reported varying step heights for FLG supported on silicon oxide [23, 15, 16]. This variation in the thickness of the single graphene layers may be attributed to the change in the tip – sample interaction as the tapping tip scans over the surface. Observations of distortions in the thickness of nanoparticles, measured with TAFM, are well known. Anomalous nanoparticle height measurements, dependent on the free amplitude of the cantilever and material properties of the sample, were reported earlier [17, 18, 19].

* Corresponding author. Fax: +36-1-3922226
email address: nemes@mfa.kfki.hu, URL: *www.nanotechnology.hu* (Peter Nemes-Incze)

It is generally accepted that folded regions in the graphene give the most reliable measurement of thickness [3], however such folded regions are not always available in every experiment. Furthermore, samples of good quality should not contain such regions, leaving no option, but to check the thickness relative to the oxide surface. Recently Raman scattering [20, 21, 22] and Rayleigh scattering [23] have been shown to be useful tools in determining the number of graphene layers in a given sample, and other optical techniques are showing promise [13, 14, 24]. However AFM is still being used frequently to determine, or confirm the thickness of FLG layers obtained by other techniques [23, 25]. Our investigations shed light on some of the precautions that must be considered, when estimating FLG film thicknesses using TAFM.

The fact that the thickness of graphene films measured by different groups has a certain deviation suggests that the data obtained are dependent on either the measurement conditions, sample preparations procedures or other laboratory conditions. In this paper, we investigate the dependence on scanning parameters of the measured thickness of one and multilayer graphene films on silicon oxide surfaces, using TAFM. We show that the instrumental offset in the thickness is greatly influenced by the free amplitude of the tapping cantilever. Differences of as much as 1 nm can be observed in the measured height of the very same graphene platelet. We have compared Raman spectroscopy, TAFM and contact mode AFM (CAFM) measurements on selected FLG crystals. We present experimental evidence that the most important factors in distorting topography data, in TAFM may be the improperly chosen free amplitude of the cantilever and the amplitude setpoint.

**Experimental**

Graphene samples have been prepared by mechanical exfoliation [3]. HOPG (SPI-1 grade, purchased from SPI Supplies) was rubbed against a silicon wafer covered by a 300 nm layer of silicon oxide. Graphene and few layer graphite crystals were identified using optical microscopy according to the procedure described in [2].

A Multimode Nanoscope SPM, from Veeco with a IIIa controller, was used in tapping mode and contact mode to characterise the FLG samples, under ambient conditions. Silicon scanning tips used in tapping mode were purchased from Nanosensors (model: PPP-NCHR), with tip radiuses smaller than 10 nm (force constant ~42 N/m, resonance frequency in the range of 300 kHz). The cantilever drive frequency was chosen in such a way as to be 5% smaller then the resonance frequency. The free amplitudes of the TAFM tips used were determined from amplitude – distance curves. Raman spectra were recorded on selected graphene films, using a Renishaw 1000 MB Raman microscope. The excitation source was the 488 nm line of an $Ar^+$ laser with incident power in the mW range in order to avoid excessive heating of the sample, using a laser spot with a diameter of 2 μm.

**Results and Discussion**

TAFM images are not purely topographic, but depend on the material properties of the sample and the interaction forces between the tip and the sample. Measurements with TAFM, at constant amplitude show a surface of constant damping of the cantilever oscillation.

We have studied various FLG crystals, by TAFM. During these measurements we have noticed inconsistencies in the measured thickness, relative to the oxide layer. That is to say, the thickness measured over a given FLG crystal was not always the same and in some cases we have even observed random switching of the FLG thickness in the same image. It is known that the free amplitude and setpoint of the TAFM cantilever can have significant influence on the imaged topography [17, 18, 19, 26]. As the amplitude of the cantilever is the main signal, based on which, the topography is mapped, we have investigated the influence of the cantilever free amplitude on the measured thickness of FLG crystals. In doing so, we have measured the thickness of various flakes, using a range of free amplitude settings, keeping the setpoint of the cantilever amplitude at a constant value. The results of such a measurement can be seen in Figure 1. Two FLG crystals were measured simultaneously, one overlapping the other. The free amplitude was varied from 16 nm to 30 nm. For each free amplitude setting a complete AFM image was acquired and the step heights in three regions were evaluated (marked by white squares): crystal C1 – oxide; crystal C2 – oxide and C2 overlapping C1. Starting with the 16 nm free amplitude and keeping the setpoint constant, we have observed that at 26 nm free amplitude, the thickness measured on top of the oxide surface decreases almost instantly, by about 0.8–1 nm. However, the thickness measured at the overlapping region of FLG C2 (green triangles) stays constant. This shows, in accordance with reports from literature that a more reliable measure of thickness is the step height relative to another graphite substrate. The effect described here was checked on various FLG crystals, using different scanning tips. In each case, the effect could be observed, to a greater or minor degree, with deviations in the thickness measured at low free amplitudes.

The presence of two “stable” thickness values hints at the existence of a bistability in the measurement system. To further investigate the phenomenon, we have measured amplitude – distance (AD) curves on a FLG surface and the neighbouring oxide substrate, using a range of free amplitudes. The curves were obtained by reducing the tip sample distance from a value larger than the free amplitude to a minimal separation, where the amplitude was reduced to about 10% of the free amplitude. The amplitude was not reduced to zero because in this manner the reproducibility of the AD curves was poor [27].

A typical AD curve is plotted on Figure 2a., recorded on an FLG surface, at 25.8 nm free amplitude. The striking feature of the amplitude curve is that at the amplitude value of 16 nm a jump can be observed. In this region, two different piezo displacement values correspond to the same amplitude, the difference being about 1 nm. This is important because the feedback electronics of the AFM works correctly only for a linear signal. If the measurement setpoint is selected in such a way as to coincide with the jump in amplitude, the feedback electronics may produce random switching from one displacement value to the other [26]. Since the height signal is derived from the piezo displacement signal, random switching in height occurs. This behaviour is presented in Figure 2b-c on a FLG film. In one case Fig. 2b. the imaging is stable on silicon oxide, while in Fig.2c. stable imaging is achieved over the FLG.

Changes in topography of such a magnitude (~1 nm) have been reported previously by Kühle et al. [19] on Cu clusters supported on a silicon oxide substrate. The origin of this change in topography, as reported by the authors, is a jump in the amplitude response of the cantilever, with changing tip – sample separation, as seen on Fig. 2a. Anczykowski et al, using time resolved numerical simulation of the tapping tip [28] pointed out that the jump in amplitude marks a change in the sign of the tip sample interaction force. When the tip starts to approach the sample, the amplitude decreases linearly. In this regime, long range attractive forces are responsible for the oscillation damping. At a certain tip – sample separation a jump occurs in the amplitude (see Fig. 2a). This jump marks the onset of a region where, with further decreasing tip – sample distance, both long range attractive and short range repulsive forces act on the tip, ie. the tip is in hard mechanical contact with the sample. After the jump, the damping of the oscillation increases further, but this time net repulsive forces characterise the tip sample interaction and the contact time of the tip also produces a jump [29]

In the following paragraphs we will discuss the effect of this behaviour on topography, by the example of a measurement on a FLG flake. In Figure 3. we have plotted the AD curves on the FLG and oxide surface at three different free amplitude settings: 24 nm, 26 nm and 28.5 nm. The setpoint value is 15 nm in all cases. We can observe the presence of net attractive and net repulsive regimes of interaction on both surfaces, with the effect being more pronounced on the FLG. One important characteristic is worth pointing out: at 24 nm amplitude the measurement setpoint is in the net attractive regime for both the oxide and FLG surfaces. When we increase the free amplitude the curve shifts and at 26 nm the setpoint is at the instability point on the graphite. It is exactly at this free amplitude that the TAFM image in Figure 2b was acquired. Because of the presence of two piezo displacement values for a certain amplitude the feedback electronics can not distinguish between these two, as it needs a linear signal to work with. Therefore random switching of the step height on the FLG surface can be seen, similar to the effect observed by Kühle et al. [19, 26]. Increasing the free amplitude further, the setpoint on both the oxide and FLG surface will be in the net repulsive region. Plotting the step height dependence on the FLG as a function of free amplitude we obtain the graph in Fig. 4. The plot shows a steep decrease in the step height at around 26 nm amplitude, from 4.5 nm to 2.25 nm. This falls within the range of the piezo displacement jump, when transition occurs into the region where repulsive interactions become dominant (see Fig. 2a). Considering the information on Figure 1c that at free amplitudes of 26 nm and higher the step height measured above the oxide and flake C1 correspond, we can say that a more precise measure of the step height can be obtained, when measuring in the repulsive regime on both oxide and FLG.

To crosscheck our data and to support the claim that the measured height (thickness) of FLG crystals is influenced by the selected free amplitude, we have performed Raman scattering experiments on graphene and FLG crystals having different numbers of layers. FLG flakes of 1-5 layers were proven to exhibit characteristic Raman signatures [21].

We have carefully investigated graphene and FLG flakes (2, 3, 5, 10 layers) by TAFM at different free amplitude settings and Raman spectroscopy, using a 488 nm laser. The Raman spectra of single, bilayer and trilayer graphite are displayed in Figure 5. The distinct characteristics [21] of the 2D peak of graphene and bilayer graphite at ~2700 $cm^{-1}$ are clearly identifiable. The Raman spectra of trilayer graphite are also clearly distinguishable from the bulk signal. TAFM images and linecuts of the same FLG flakes used to acquire the Raman spectra are shown in Figure 6. Topography images acquired in the repulsive regime, at high free amplitude settings on both FLG and oxide surfaces fit the Raman data well. Furthermore, the images measured using low free amplitudes show a much larger thickness value (see Fig. 6). Where possible, we measured the thickness of folded regions (like in the case of Fig. 1a). Such observations further support the claim that in order to gain reliable thickness data, one needs to be using a setpoint where the tip scans in the net repulsive regime on both the oxide and FLG surfaces. It is worth mentioning that true, single layer graphene crystals were frequently measured to be around 1 nm thick using free amplitudes below the sharp drop in thickness.

In the example of the FLG layer in Fig 2b, further decreasing the free amplitude shifts the setpoint to a region where the damping is of attractive type on both surfaces, this time with the instabilities in topography appearing on the oxide (see Fig. 2c). In this case a thickness of 3.5 nm is measured (see Fig. 4), which is still different from the more precise thickness of 2.3 nm (about 7 layers). This goes to show that the damping of the amplitude due to attractive forces is of a different value for the two surfaces. Attractive forces acting on the sample have various components: electrostatic, Van der Waals, capillary or chemical forces. On mica and graphite surfaces one of the strongest contributions to the attractive force comes from the capillary forces, as demonstrated by Ouyang et al. [30]. This comes as no surprise, since under ambient conditions a thin water layer is present on most surfaces. Due to the strong hysteretic nature of the capillary force, its contribution to oscillation damping is large [27]. According to our observations, at small free amplitude values, the tip does not even enter the repulsive regime and no jump in amplitude will occur. This is in accordance with the measurements and simulation studies of Zitzler et al. [27], who have also demonstrated experimentally that the free amplitude at which a jump in the AD curves occurs is strongly dependent on the ambient relative humidity, further proof of the fact that capillary forces have a key role to play in cantilever damping.

On surfaces with changing material properties, due to differences in wettability, or more generally speaking gradients in the attractive forces, TAFM measurements of topography are not reliable [27]. Therefore, as seen in our experimental results, it is advised that the TAFM measurements on graphene be carried out with great care and measurement setpoints be chosen in the repulsive interaction regime, where the damping of the cantilever is largely due to the topography of the sample. Before measurements amplitude distance curves should be acquired on both the graphene and oxide surface and the measurement setpoint chosen accordingly.

In our measurements of FLG flakes, the difference in height comes from the fact that the repulsive region sets in at different free amplitude for the graphite and oxide. As the amplitude setpoint crosses the jump in the AD curve unstable imaging is observed on both the FLG and oxide. However, the shift of the critical amplitude at which the repulsive regime sets in and its dependence on the nature of the attractive forces is still not fully understood. Further research in this direction is on the way.

We have also performed CAFM measurements on our FLG samples, which further confirm our findings. However, we found that the difference in lateral forces on the FLG and support can introduce deviations in the thickness measured using contact mode. This is illustrated by Figure 7, where we present TAFM and CAFM measurements on the same FLG flake. The thickness of 2 nm (6 layers) measured at high free amplitudes in tapping mode, correlates reasonably well with the thickness measured using contact mode. However, we have observed a difference in the thickness measured by CAFM when changing the direction of scanning (no such effect was observed using TAFM). This suggests that differences in lateral forces on the FLG and oxide surfaces (for example friction) play a non negligible role in influencing the CAFM cantilever bending, resulting in differences in measured thickness. Such forces are negligible in TAFM.

## Conclusions

While performing TAFM measurements on graphene and FLG flakes on silicon oxide substrates, special care needs to be taken to obtain more precise flake thickness data. The change in FLG thickness described in this paper, is in the order of 1 nm, which is a very large error, when working with topography changes of less then one nanometre.

Where possible, the measurements should be performed in the repulsive regime on both oxide and FLG surface. Either the free amplitude or measurement setpoint should be chosen in such a way as to obtain this condition. Usually the setpoint is chosen in such a way as to be as near as possible to the free amplitude in order to minimise the forces acting on the tip and sample, but as we have seen, this may not be the correct setting. We believe that the variation of the reported thickness of graphene among different research groups is largely due to the unreliability of TAFM measurements in the attractive regime.

It is worth noting here that during our measurements at high free amplitude and low setpoint settings, no damage to either the sample or tip were encountered.

Our work sheds some light on how a more precise estimate of the number of layers in a FLG crystal can be obtained. However the number of layers should be compared to data obtained by other methods as well, to support the AFM data. Furthermore additional experimental work and computer simulation needs to be done, to determine the cause of the shift in the instability point on the AD curves and the nature of the attractive forces on the sample.

**Acknowledgements**

The present work was financially supported by Hungarian Scientific Research Fund OTKA-NKTH K67793, OTKA-NKTH NI67702 and and OTKA-NKTH 67842 grants.

**Figures:**

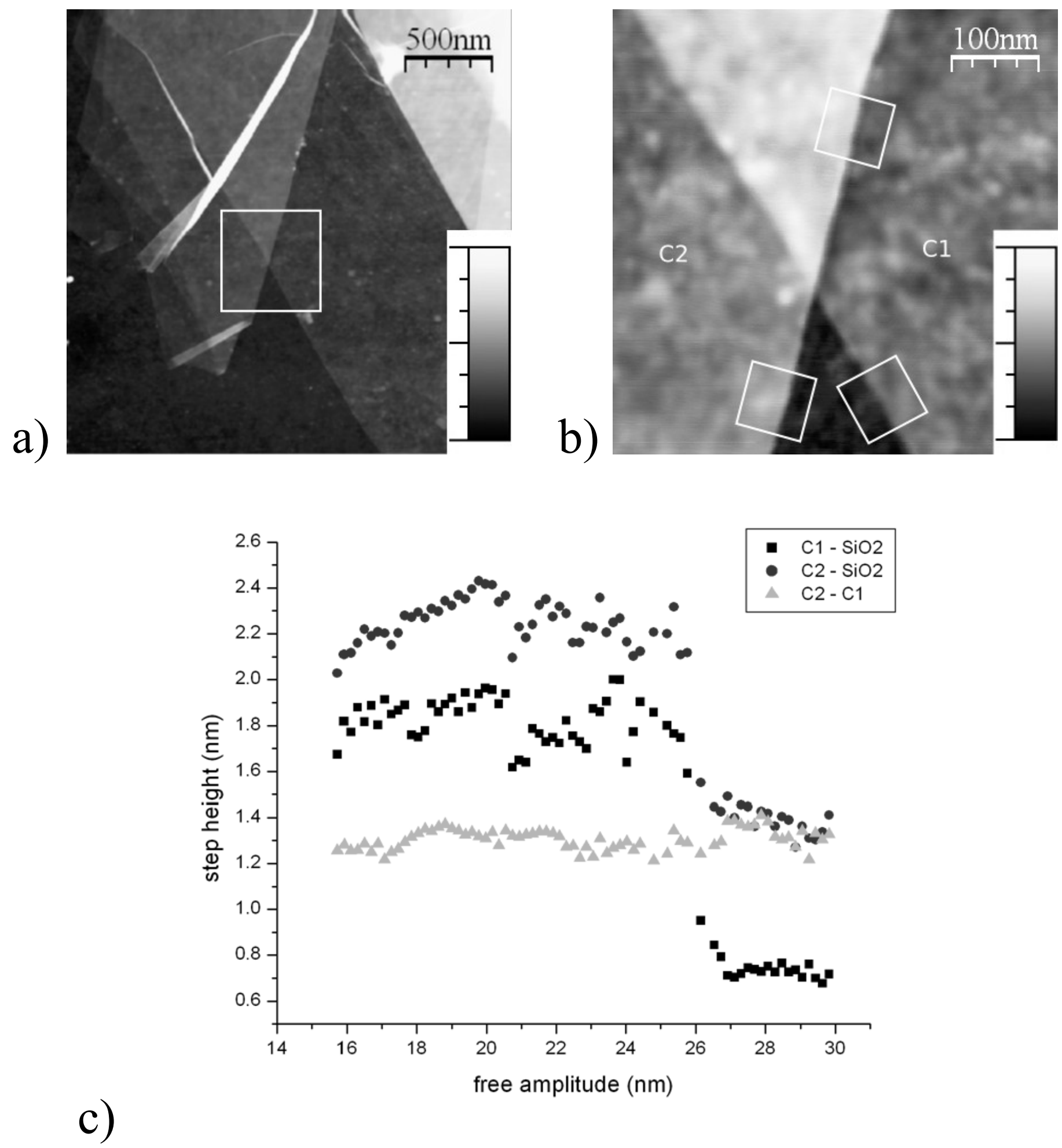

**Figure 1.** a) Image of two overlapping FLG films on $SiO_2$ (colour bar associated to height: 10 nm) b) Zoomed in region from image 1a (colour bar associated to height: 3 nm), averaged line cuts and thickness measurements were taken in the regions marked by squares, C1 and C2 representing each FLG film. In fig. c), each point represents the thickness of the crystal C1 (black squares) and C2 (red circles) with respect to the oxide substrate, as a function of free amplitude. The thickness of crystal C2 overlapping C1 with changing free amplitude is also plotted (green triangles).

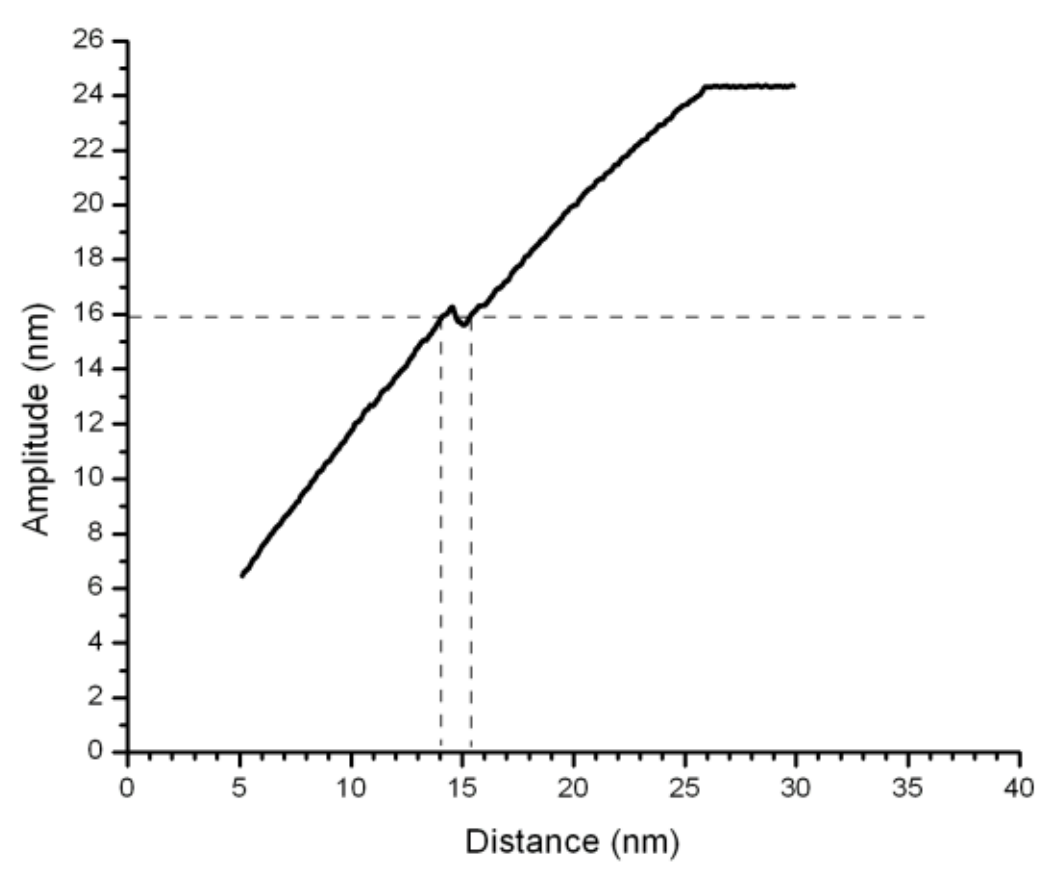

**Figure 2a.** The damping of the cantilever oscillation as

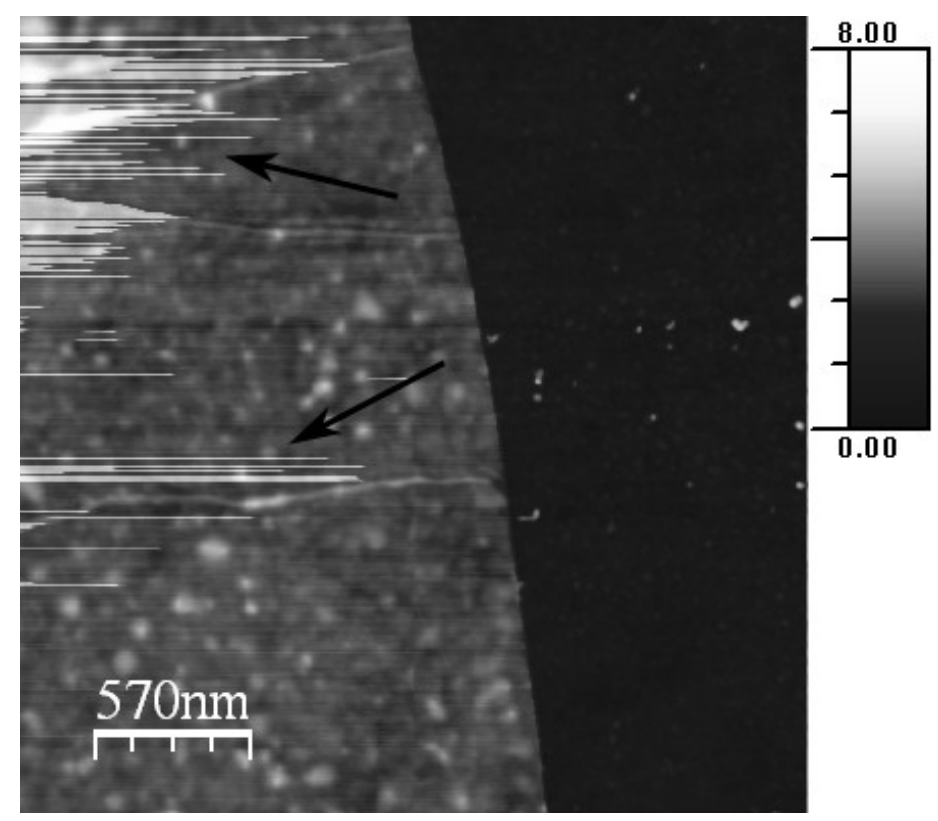

**Figure 2b.** TAFM image of a FLG flake, imaged at a setpoint of 15 nm, near the bistability point in the AD curve. Random switching from one thickness to the other

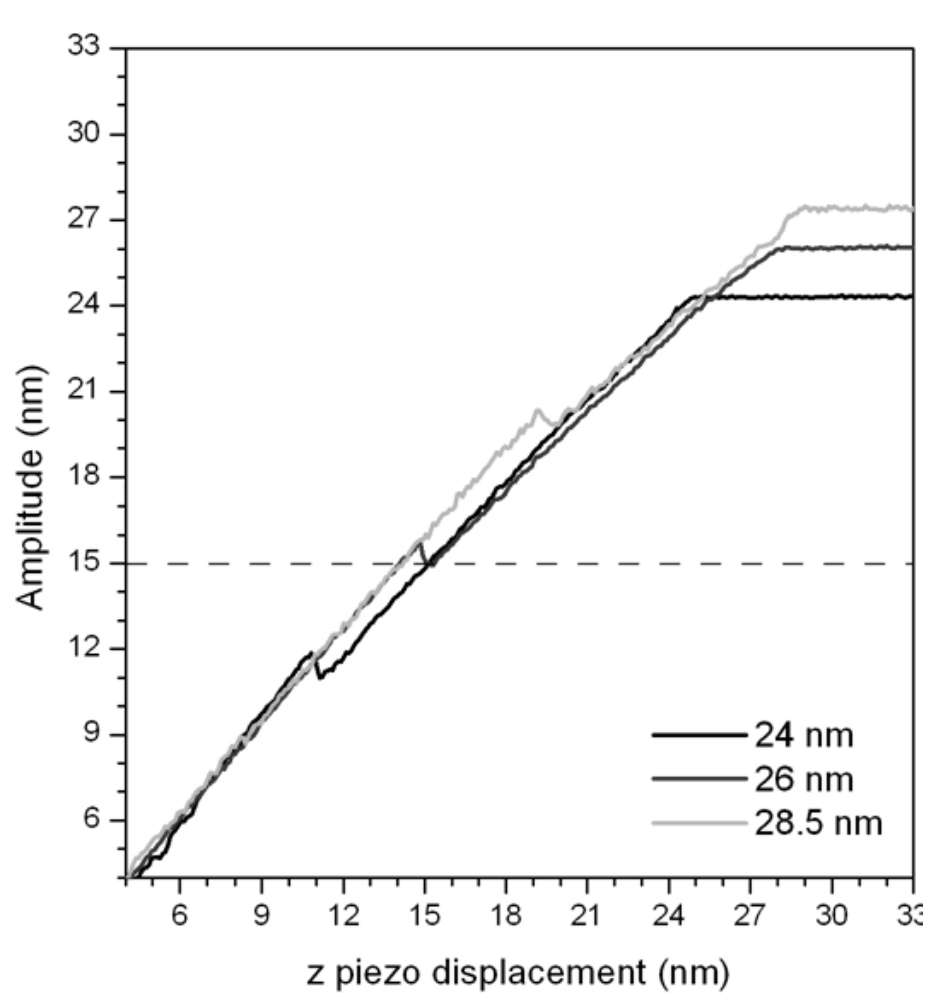

a)

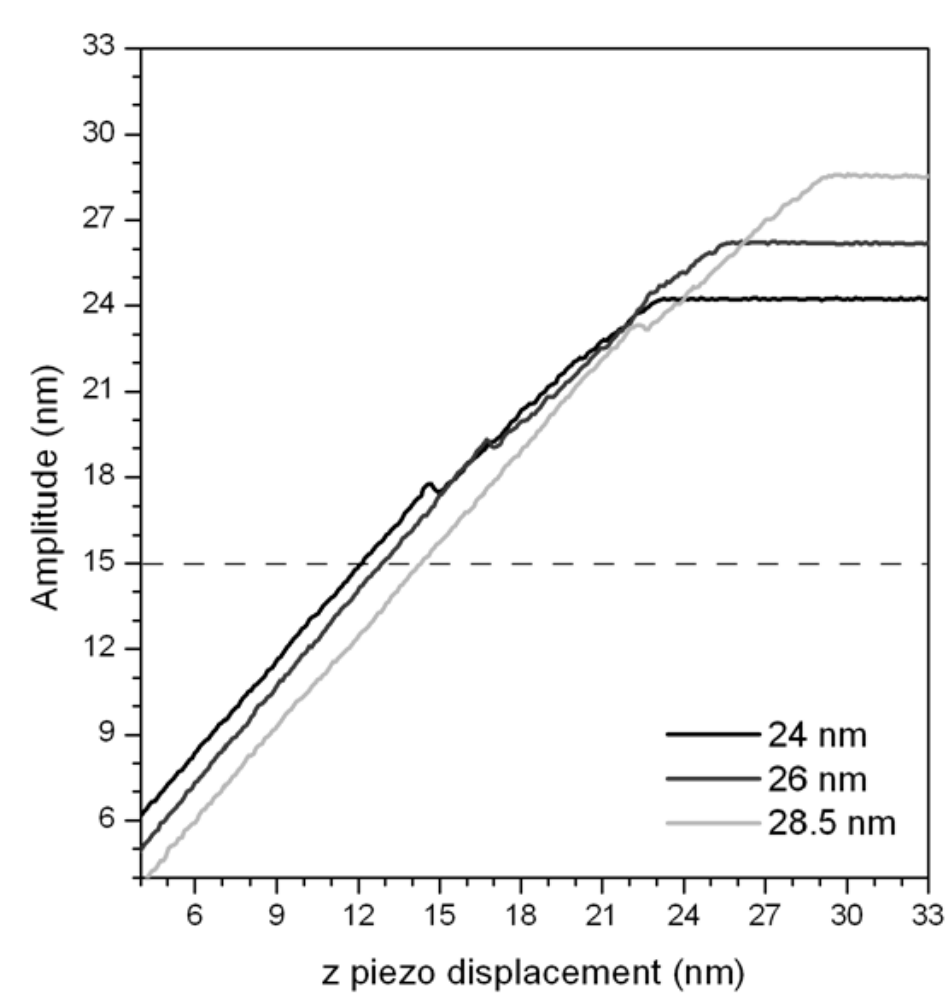

b)

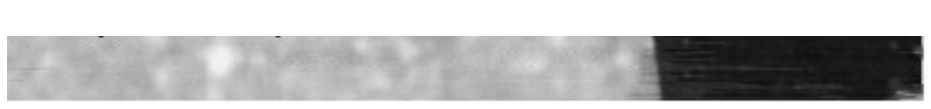

**Figure 2c.** TAFM image on the FLG flake, when

the ir

ampl

imag

arrov

**Figure 3.** AD curves on the FLG (a) and oxide (b) surface at different free amplitudes (24, 26, 28.5 nm). Dashed line at 15 nm amplitude shows the setpoint used during measurements. At 26 nm free amplitude the setpoint crosses the jump in amplitude. For the oxide surface such a crossing is experienced at free amplitudes around 21 nm.

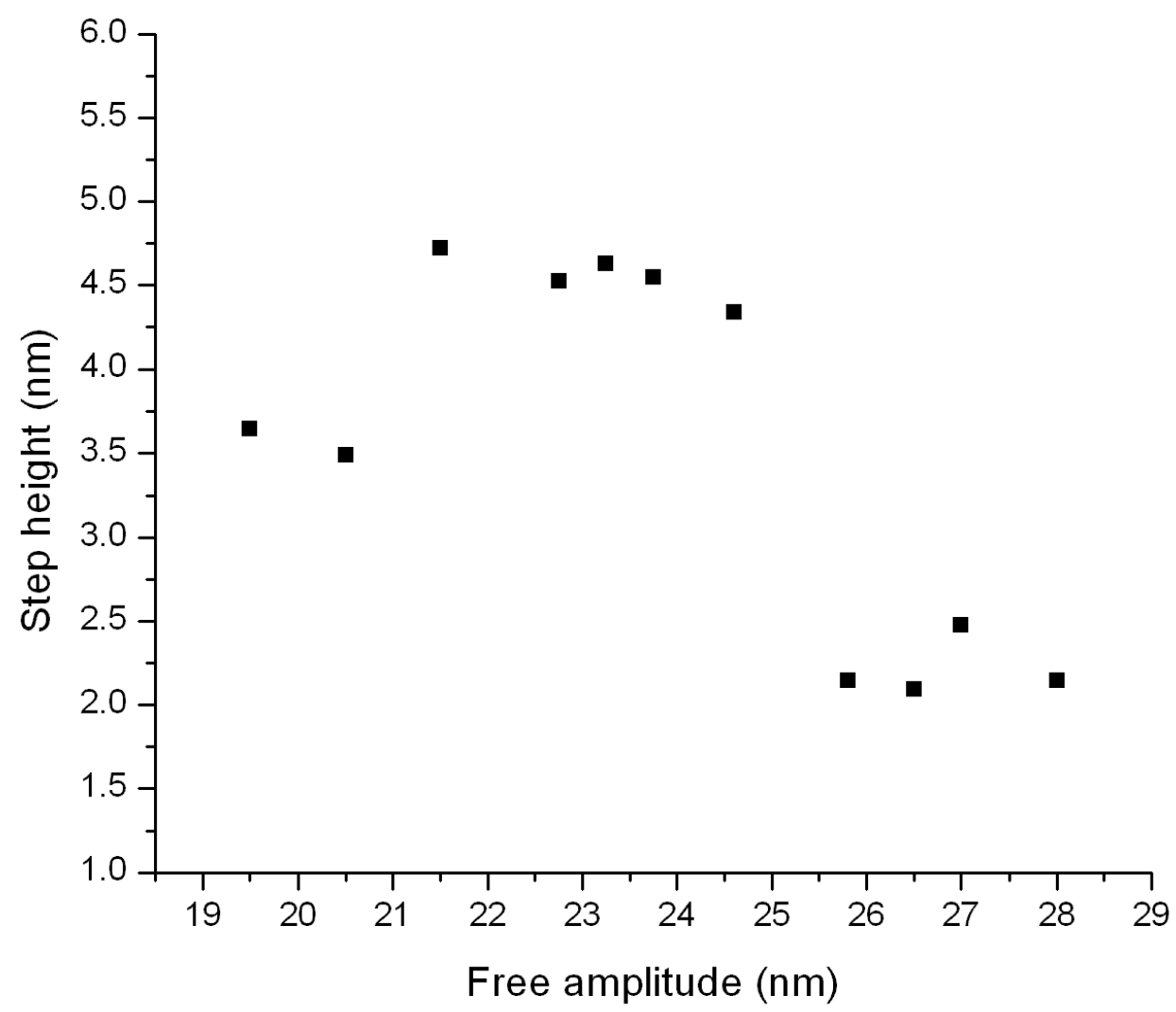

**Figure 4.** Step height as a function of free amplitude, measured on the FLG flake presented in Fig. 2b. Two jumps in height can be observed: the jump at ~25 nm marks the transition of from the attractive to the repulsive regime on FLG and the one at ~21 nm the transition from attractive to repulsive on the oxide surface.

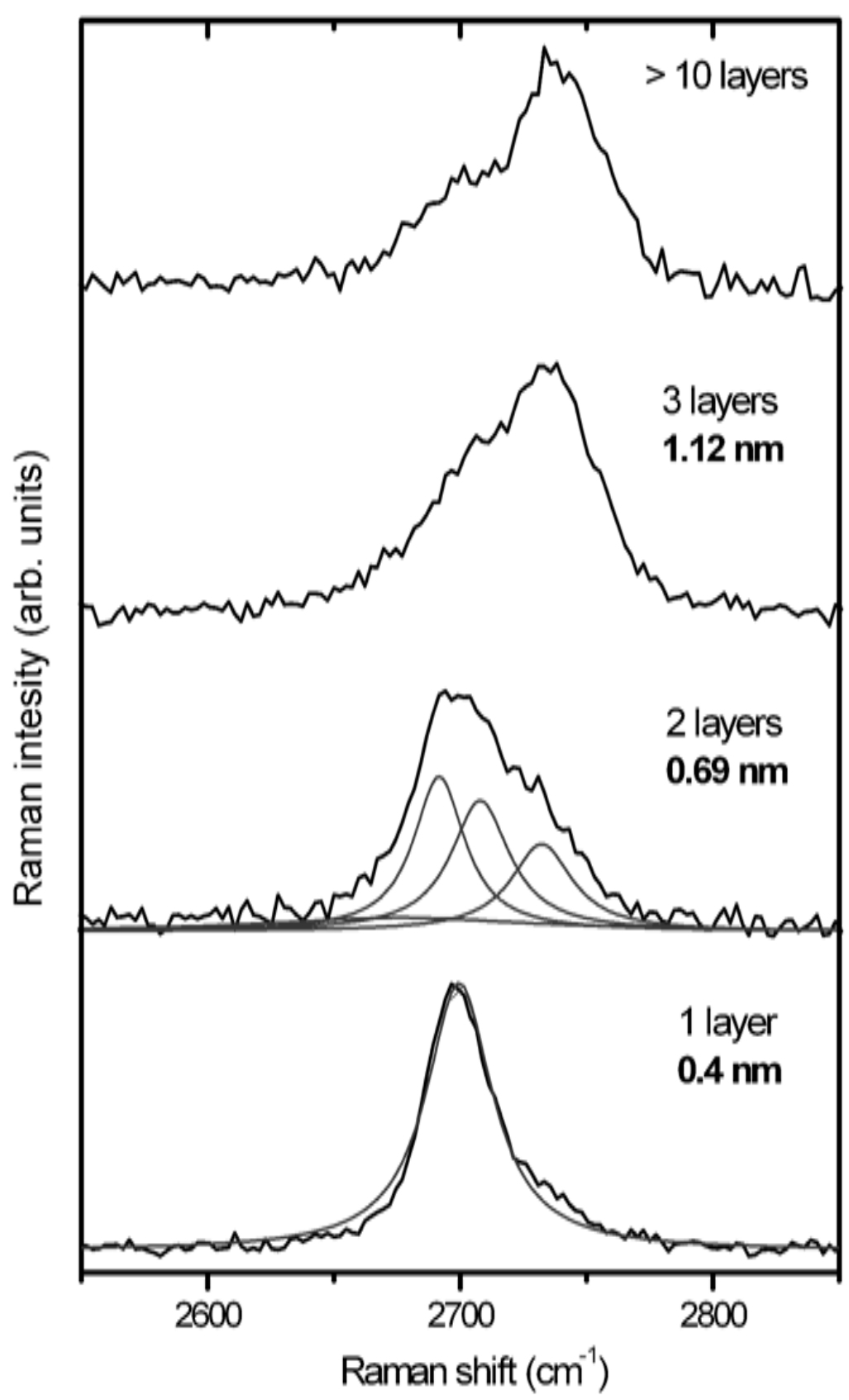

**Figure 5.** Raman spectra of graphite having 1, 2, 3 and >10 layers (scaled to have similar height of the 2D peak). For each sample we show the height measured by TAFM, at high free amplitudes. These values correlate well with Raman spectra. The four components of the 2D peak in bilayer graphite are plotted.

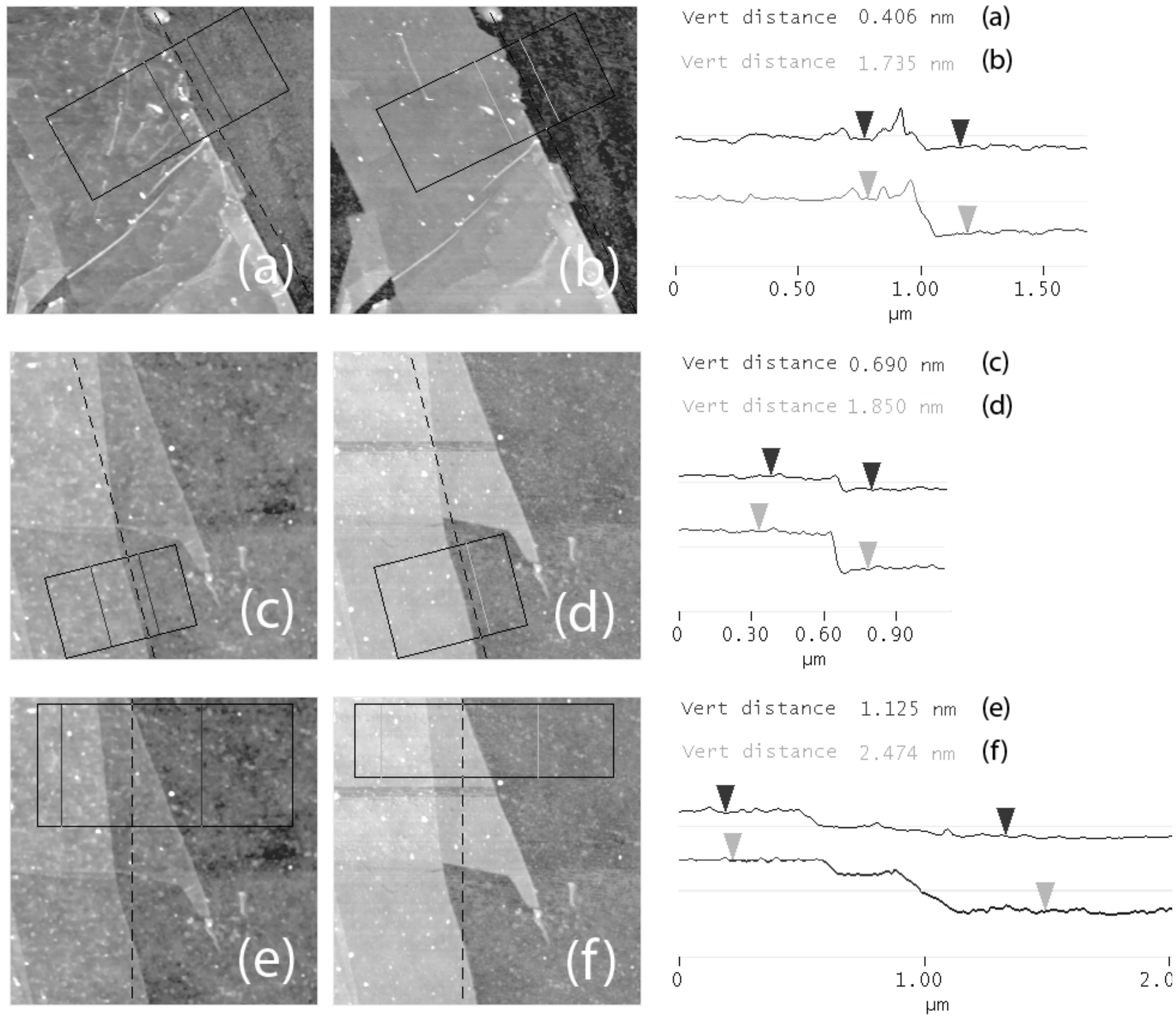

**Figure 6.** TAFM images of the regions where the Raman spectra in Fig. 5 have been acquired, for single layer (a, b); bilayer (c, d); three layer (e, f) flakes (each image is 2.5 µm x 2.5 µm). The images were acquired using a constant amplitude setpoint and two different free amplitudes, one higher (a, c, e) and another one lower (b, d, f) than the amplitude at which unstable imaging occurs on the FLG. In the first case (a, c, e) the setpoint is in the repulsive regime for both oxide and FLG, while in the latter case (b, d, f) imaging is in the attractive regime for FLG. Averaged linecuts (inside the black markers) taken on each image show the increase in thickness when measurements are not performed in the repulsive regime.

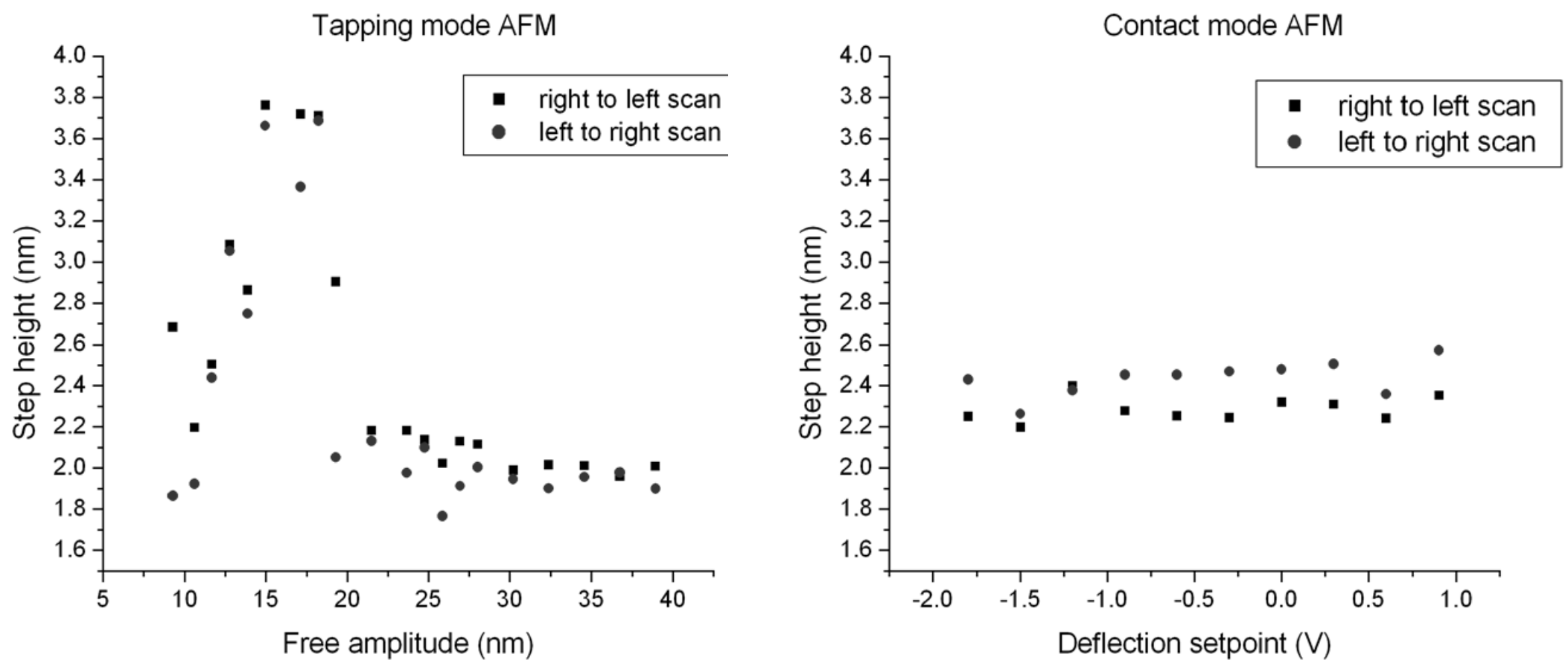

**Figure 7.** TAFM measurements of a FLG step height at various free amplitudes (left image). CAFM measurement of the same flake, using various deflections of the cantilever, ie. changing the contact force (right image). Changing the contact force does not have any effect on the step height, but the left to right and right to left scans differ by about 0.2 nm.